\documentclass[multphys,vecphys]{svmult}

\usepackage{makeidx}     
\usepackage{graphicx}    
\usepackage{multicol}    

\def\asec{\ifmmode ^{\prime\prime}\else$^{\prime\prime}$\fi}

\def\degs{\ifmmode ^{\circ}\else$^{\circ}$\fi}
\def\amin{\ifmmode ^{\prime}\else$^{\prime}$\fi}
\def\asec{\ifmmode ^{\prime\prime}\else$^{\prime\prime}$\fi}

\def\degs{\ifmmode ^{\circ}\else$^{\circ}$\fi}
\def\amin{\ifmmode ^{\prime}\else$^{\prime}$\fi}

\def\EE#1{\times 10^{#1}}

\def\cm3{\rm ~cm^{-3}}
\def\kms{\rm ~km~s^{-1}}

\def\Ti44{M(^{44}{\rm Ti})}

\def\Msunyr{~{\rm M}_\odot~{\rm yr}^{-1}}
\def\Mdot{\dot M}
\def\EE#1{\times 10^{#1}}
\def\cm3{\rm ~cm^{-3}}
\def\kms{\rm ~km~s^{-1}}

\def\lsim{\!\!\!\phantom{\le}\smash{\buildrel{}\over
  {\lower2.5dd\hbox{$\buildrel{\lower2dd\hbox{$\displaystyle<$}}\over
                               \sim$}}}\,\,}
\def\gsim{\!\!\!\phantom{\ge}\smash{\buildrel{}\over
  {\lower2.5dd\hbox{$\buildrel{\lower2dd\hbox{$\displaystyle>$}}\over
                               \sim$}}}\,\,}

\makeindex             

\begin{document}

\title*{High-resolution optical studies of nearby Type Ia supernovae}
\author{Peter Lundqvist\inst{1}
\and Seppo Mattila\inst{1}
\and Jesper Sollerman\inst{1}
\and E. Baron\inst{2}
\and Pascale Ehrenfreund\inst{3}
\and Claes Fransson\inst{1}
\and Bruno Leibundgut\inst{4}
\and Ken'ichi Nomoto\inst{5}
}
\authorrunning{P.~Lundqvist et al.} 

\institute{Stockholm Observatory, AlbaNova, Department of Astronomy,
SE-106 91 Stockholm, Sweden
\texttt{peter@astro.su.se}
\and
Department of Physics and Astronomy, University of Oklahoma,
Norman OK 73019-0225, USA
\and
Leiden Observatory, P.O. Box 9513, 2300 RA Leiden, Netherlands
\and
European Southern Observatory, Karl-Schwarzschild-Strasse 2,
D-85748 Garching bei M\"unchen, Germany
\and
Department of Astronomy and the Research Center for the
Early Universe, University of Tokyo, Tokyo 113-0033, Japan
}
%
%
\maketitle


\begin{abstract}
Since April 2000, a program using the ESO/VLT/UVES\footnotemark{} to search 
for \footnotetext{Based on observations collected at the European Southern
Observatory, Paranal, Chile. Program ID 67.D-0227(A).}
early circumstellar signatures from nearby supernovae has been
conducted.\footnotemark{}
\footnotetext{See http://www.astro.su.se/$\sim$peter/sntoo.html. In 
collaboration with the SN~Ia RTN team (PI: W. Hillebrandt), see
http://www.mpa-garching.mpg.de/$\sim$rtn}
Until now, two Type Ia supernovae have been observed, SNe~2000cx and 2001el. 
Here we report on preliminary results for SNe~2001el, and we discuss how the
observations can be used together with detailed modeling to derive an upper
limit on the putative wind from the progenitor system. For a hydrogen-rich
wind with velocity 10$\kms$, the mass loss rate for the progenitor system
of SN~2001el is $\Mdot \lsim 1\EE{-5} \Msunyr$.
\end{abstract}

\section{Introduction}
\label{sec:1}
The origin of Type Ia supernovae (SNe~Ia) is still unclear. 
Branch et al.~\cite{bly95} list possible types of systems, and argue that
the most likely system is a C-O white dwarf which accretes matter from the
companion, either through Roche lobe overflow, or as a merger with another
C-O white dwarf. We need methods to discriminate between the 
possible progenitor scenarios. In non-merging scenarios a wind from the
companion star is expected. If the wind is ionized and dense enough, it
could reveal itself in the form of narrow lines before being overtaken by
the supernova blast wave, just as in narrow-line core-collapse supernovae,
SNe~IIn. For a hydrogen-rich wind, H$\alpha$ would be emitted, and if
helium dominates He~I~$\lambda\lambda$5876, 10830 and He~II~$\lambda$4686
may be prominent.

A pioneering observational and theoretical study of circumstellar line
emission in SNe~Ia was that of SN~1994D~\cite{cl96,lc97}. The spectrum was 
obtained $\sim 6.5$ days after explosion with a spectral resolution 
of $\sim 30 \kms$, and covered H$\alpha$. The analysis of the non-detection 
involved full time-dependent photoionization calculations to estimate the 
narrow H$\alpha$ emission from the tentative wind. The
analysis in~\cite{lc97} gave an upper limit on the mass loss from the 
progenitor system of $\Mdot \lsim 2.5\EE{-5} \Msunyr$, assuming cosmic 
abundances for the wind, and a wind speed of $v_{w} = 10 \kms$. More 
recently, Lundqvist et al.~\cite{l03} studied SN~2000cx with the 
Ultraviolet and Visual Echelle Spectrograph (UVES) on ESO's Very Large 
Telescope (VLT). The spectral resolution of UVES is $\approx 50,000$, 
i.e., $6~\kms$ in the region of Na~I~D and H$\alpha$. The preliminary upper 
limit on the mass loss rate from the progenitor system 
is $\Mdot \lsim 9\EE{-6} \Msunyr$ for $v_{w} = 10 \kms$. Here we report
on the second of our SNe~Ia observed with VLT/UVES, SN~2001el.

%

\section{Observations and Results}
\label{sec:2}
SN~2001el was discovered on September 17.1 2001~\cite{mon01}.
It was situated 9" west and 15" north of the nucleus of the nearby
(recession velocity v$_{rec}$ = 1164 km~s$^{-1}$) galaxy NGC 1448.
SN~2001el was observed as part of our VLT/UVES Target-of-Opportunity (ToO)
program on September 21. This spectrum had a complete wavelength coverage 
between $3260-10600$~\AA\ and allowed a classification of the 
supernova as a Type Ia observed well before maximum~\cite{sol01}.
SN 2001el was also observed on September 26, and then
revisited for the last time on September 28.
Our UVES observations were thus obtained 9, 4 and 2 days before the
supernova B-band maximum light~\cite{kri03}.

\begin{figure}
\centering
\includegraphics[height=9cm, angle=270]{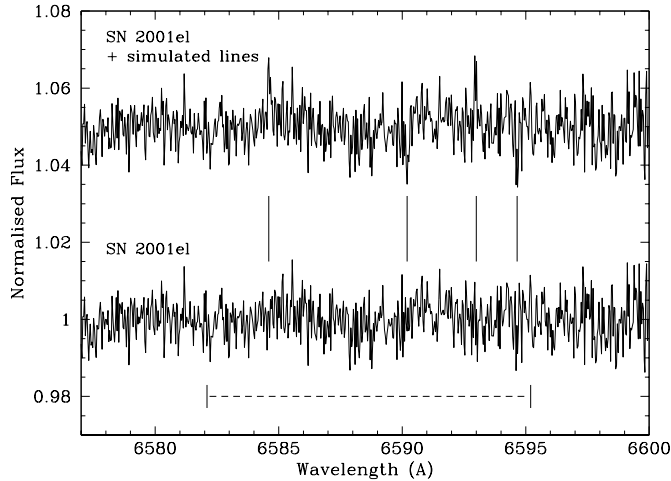}
%
%
\caption{
Normalized third epoch (Sep 28.3 UT) spectrum of SN~2001el showing the
region around the expected wavelength range of H$\alpha$ (marked with a
horizontal dashed line). The lower and uppermost plots, respectively, show
the SN spectrum before and after including simulated 3$\sigma$ features with
a Gaussian profile and FWHM of 6 km~s$^{-1}$. The locations of these
features at 6584.6~\AA\ (emission), 6590.2~\AA\ (absorption), 6593.0~\AA\
(emission), and 6594.7~\AA\ (absorption) are marked with vertical lines. The
spectra have been rebinned by a factor of 2, and thus have a pixel size of
$\sim$1.6 km~s$^{-1}$. The uppermost spectrum has been moved upward for
clarity.
}
\label{fig1}
\end{figure}

The lines expected from a hydrogen- or helium-rich companion wind
were sought for in the SN spectra at the three epochs of
observation at the expected range of wavelengths assuming $v_{rec} =
(1180\pm300)$~km~s$^{-1}$ for the supernova. No such lines were
detected, and therefore upper limits for the emission line fluxes and
absorption line equivalent widths were derived. This was done by fitting 
Gaussian profiles with fixed FWHMs at fixed positions within the expected 
range of wavelengths in the unbinned spectra. The 1$\sigma$ levels for the 
emission line fluxes and absorption line equivalent widths were then 
obtained by requiring 68.3~$\%$ of the measured line fluxes (or equivalent 
widths) to be below this level. Adopting an extinction of A$_{V} = 0.83$, 
a distance of $d = 18$ Mpc for SN 2001el and assuming that the line has a 
FWHM of $10\kms$ (see Mattila et al. in prep.), we obtain an upper limit 
for the H$\alpha$ narrow line-luminosity 
of $L_{{\rm H}\alpha} = 3.2\times10^{36}$ ergs~s$^{-1}$ in our third epoch
(Sep 28.3 UT) spectrum. To reach this limit, we performed a careful backgroud 
subtraction of the host galaxy emission. This was done manually outside the 
UVES pipeline.

Figure~1 shows examples of artificially introduced unresolved emission and 
absorption lines at the 3$\sigma$ level. These lines with FWHM~$=6$~km~s$^{-1}$
were simulated at the expected wavelength range of H${\alpha}$ in the unbinned
third epoch UVES spectrum. The spectrum was then rebinned by
a factor of 2 resulting in a pixel size of $\sim$$\frac{1}{4}$ $\times$ FWHM
in the rebinned spectrum. The simulation parameter for the line peak was
selected for each line such that the measured apparent flux of the line
corresponds to the 3$\sigma$ level. No features with fluxes (or equivalent
widths for the absorption lines) as high as the ones of the simulated lines
are apparent in the real data.

\section{Modeling and Discussion}
\label{sec:3}
We have modeled the line emission in a way similar to
what was done in~\cite{cl96,lc97,l03}. We assume that the supernova ejecta 
have a density profile of $\rho_{ej} \propto r^{-7}$, and that the ejecta 
interact with the wind of a binary companion which has a density profile 
of $\rho_{w} \propto r^{-2}$. The power-law density distributions
makes it possible to use similarity solutions for the expansion and structure
of the interaction region~\cite{chev82a}.

\begin{figure}[ht]
\begin{center}
\includegraphics[height=7cm, angle=180]{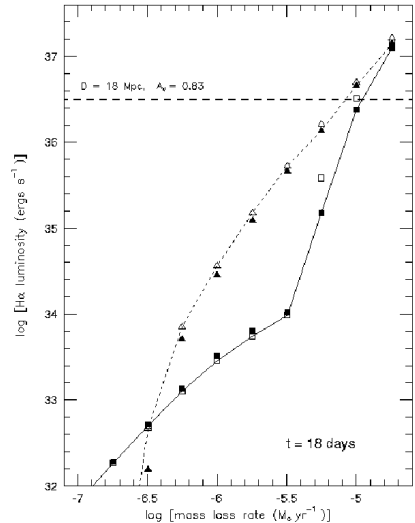}
\end{center}
\caption{
H$\alpha$ luminosity at 18 days after the explosion as a function of mass loss 
rate, assuming $v_{w} = 10 \kms$ and cosmic abundances. Triangles show models 
for which ionizing radiation is only produced by the reverse shock, while in 
models marked by squares we have also included the photospheric emission 
from the model w7jzl155.ph~\cite{blisor01}. Filled symbols are for temperature
equipartition between electrons and ions behind the reverse shock, whereas
for open symbols the electron temperature is $\frac{1}{2}$ $\times$ the 
reverse shock temperature. 
The observed limit for SN~2001el, `18~Mpc, A$_V = 0.83$', is also shown.
}
\label{fig2}
\end{figure}

Calculations are started at $t_0=1.0$~day, and
at this epoch we assume that the maximum velocity of the ejecta
is $V_{ej} = 4.5\EE4 \kms$.
At 1 day, the velocities of the circumstellar shock and the reverse shock
going into the ejecta are $V_{s} \sim 4.5\EE4 \kms$
and $V_{rev} \sim 1.1\EE4 \kms$, respectively.
The ionizing radiation from the interaction region consists of
free-free emission from the shocked ejecta and circumstellar gas, and
photospheric radiation~\cite{blisor00,blisor01,l03} Comptonized by hot 
electrons in the shocked gas. To model the time dependent ionization and 
temperature structure of the unshocked circumstellar gas we use an updated 
version of the code used in~\cite{cl96,lf96,l03}. Models were made for values
of $\Mdot$ in the range $(1-300)\EE{-7} \Msunyr$, assuming $v_{w} = 10 \kms$.
For low mass loss rates, the photospheric radiation dominates
the ionization, and its soft photons heat the wind to temperatures in the
range $(0.5-5)\EE{4}$~K, whereas for $\Mdot \gsim 5\EE{-6} \Msunyr$, the
ionizing radiation from the reverse shock becomes more important, heating
the wind close to the shock to $\gsim7\EE{4}$~K. 
Figure~2 shows the line emission produced in models with different
wind densities. Using our derived limits 
on the H$\alpha$ luminosity for SN~2001el at the third epoch, we obtain (for 
cosmic abundances and $v_{w} = 10 \kms$) a wind density described 
by $\Mdot \sim 1\EE{-5} \Msunyr$, i.e., similar to our limit for 
SN~2000cx~\cite{l03}. Such a low limit for the circumstellar gas, as 
well as other similar limits in the radio and X-rays for other SNe~Ia 
(cf. Ref.~\cite{l03}), contrasts the recently reported strong Balmer line 
emission from the SN~Ia 2002ic by Hamuy et al.~\cite{ham03}. The 
observations of SN~2002ic have been used both in favor~\cite{liv03} and 
against~\cite{cy03} a merger scenario. While the true nature of SN~2002ic 
is still under debate, detection of early, faint circumstellar line 
emission in a SN~Ia, with temporal variation in the line profiles, would 
be stronger evidence for a non-merger scenario. 
Finally, we note that the mass loss rate in symbiotic systems is in the 
range $10^{-7} - 2 \times 10^{-5}\Msunyr$~\cite{sea93}. Our results do not
support that systems at the upper end of this interval are progenitors
of normal SNe~Ia.

\smallskip
\smallskip
This work was supported by the
Swedish Research Council, the Swedish Space Board, the Royal Swedish
Academy of Sciences, and the EU RTN program RTN2-2001-00037. PL is a 
Reserach Fellow at the Royal Swedish Academy supported by a grant from the 
Wallenberg Foundation.



%
\index{paragraph}
%
%
%
%
%
%
%
%
%
%
%
%
%
%
%

%
%



\printindex
\end{document}